\documentclass[twocolumn,
prl
,superscriptaddress
]{revtex4}

\usepackage{graphicx}
\usepackage{amssymb}
\usepackage{amsmath}
\usepackage{subfigure, psfrag}
\usepackage{color}
\usepackage{comment}

\newcommand{\ket}[1]{\left|{#1}\right.\rangle}

\begin{document}


\title{Dipolar gases in coupled one-dimensional lattices}

\author{Marianne Bauer}
\affiliation{Cavendish Laboratory, JJ Thomson Avenue, Cambridge,
 CB3 0HE, United Kingdom} %

\author{Meera M. Parish}
\affiliation{Cavendish Laboratory, JJ Thomson Avenue, Cambridge,
 CB3 0HE, United Kingdom} %
\affiliation{London Centre for Nanotechnology, Gordon Street, London, WC1H 0AH, United Kingdom}

\date{\today}

\begin{abstract} 
We consider dipolar bosons in two tubes of one-dimensional lattices, where the dipoles are aligned to be maximally repulsive and
the particle filling fraction is the same in each tube. In the classical limit of zero inter-site hopping, the particles arrange themselves into an ordered crystal for any rational filling fraction, forming a complete devil's staircase like in the single tube case. Turning on hopping within each tube then gives rise to a competition between the crystalline Mott phases and a liquid of defects or solitons. However, for the two-tube case, we find that solitons from different tubes can bind into pairs for certain topologies of the filling fraction. This provides an intriguing example of pairing that is purely driven by correlations close to a Mott insulator.
\end{abstract}

\pacs{}

\maketitle

Unconventional superconductors such as the cuprates have stimulated much research on exotic pairing phenomena in low dimensions~\cite{norman_review2011}.
Superconductivity in these materials is typically observed upon doping a half-filled Mott insulating state, and thus there has been particular interest in whether pairing can be purely driven by strong electron correlations near a Mott insulator. 
Model systems that have been used to investigate this effect are the spin ladders, where one-dimensional (1D) chains of 
electrons are coupled via electron hopping and magnetic exchange interactions~\cite{dagotto92,dagottorice96}. Here, when the ladder is doped, one can obtain pairing between holes on different chains depending on the number of chains or `legs' in the ladder~\cite{dagottorice96}.  
However, the focus thus far has been on Mott insulators derived from short-range, on-site interactions and so an intriguing and physically relevant question is how \emph{long-range} interactions will affect the physics. 

It is well known that long-range interactions in 1D 
can lead to exceptionally intricate crystalline ground states.  
This is most evident in the case of classical particles on a lattice interacting via repulsive infinite-range convex potentials~\cite{hubbard78,pokrovskyu78}. Here, in the absence of any quantum kinetic energy, the particles arrange themselves into an ordered crystal, commensurate with the underlying lattice, for \emph{any} rational filling fraction. Indeed, it can be shown that the filling fraction as a function of the particle chemical potential $\mu$ forms a complete devil's staircase, where every rational filling fraction between 0 and 1 enjoys a region of stability 
within a finite interval of $\mu$~\cite{BakBruinsma}. 
Perturbing away from the classical limit, one finds that kinetic energy destroys the complete devil's staircase, but signatures of the staircase structure remain in the form of Mott lobes~\cite{burnellpcs09}. This structure has also been predicted in recent studies of the opposite weak-coupling limit, where the lattice potential is weak~\cite{Dalmonte2010}.

\begin{figure}
\begin{center}
\includegraphics[width = 0.43\textwidth]{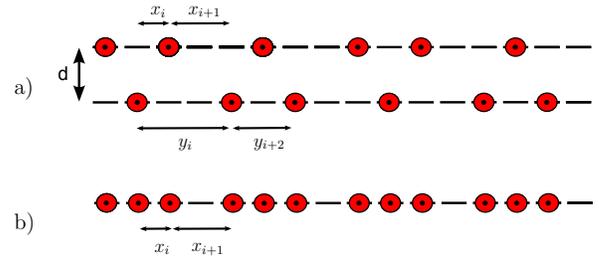}
\caption{(Color online) Arrangement of dipolar bosons (filled circles) in two 1D lattices separated by distance $d$, where the dipoles are directed out of the page. (a) Commensurate ground state for filling fraction $p/q=3/8$ in each tube. 
The ground state pattern can be determined by collapsing the two tubes onto one tube (b) and then using Hubbard's algorithm~\cite{hubbard78}  for a single tube. The distances $x_i$, $y_i$ correspond to the nearest 
and next-nearest neighbor distances, respectively, on the collapsed tube.
\label{fig:setup}}
\end{center}
\end{figure}

In this Letter, being motivated by the ladder systems, 
we extend the above problem to investigate the case of \emph{two} 1D lattices that are purely coupled by the long-range interactions. 
%
%
Such coupled 1D systems 
have recently generated much interest in the context of ultracold atomic gases owing to the possibility of confining polar molecules with long-range dipole-dipole interactions in reduced geometries~\cite{menotti,Carr2009,miranda2011}. 
Thus far, theoretical studies of tubes coupled by dipolar interactions have concentrated on the continuum limit, where there is no strong lattice potential within each tube~\cite{kollathmg08,wunschd11,Dalmonte2011,knap2011}, while here we focus on the opposite, strong-lattice limit.

We consider dipolar bosons
in two identical tubes of 1D lattices, where the 
boson filling fraction is the same in each tube and the dipoles are aligned 
so as to be maximally repulsive (see Fig.~\ref{fig:setup}). 
We focus on hard-core bosons, but our results will also apply to dipolar fermions as we note below. 
In the absence of hopping between lattice sites, we find a complete devil's staircase like in the single-tube case, but where the width of the steps $\Delta\mu$ can now depend on the intertube separation. 
For a given filling fraction $p/q$ in each tube, we find that the character of the soliton excitations (or domain walls) in the commensurate Mott phase is governed by whether $q$ is even or odd. Thus, the quantum melting of the Mott phases exhibits an unusual ``odd-even'' dependence as the intertube distance is varied. 
Crucially, when $q$ is even, we find that solitons on each tube bind together in pairs, an effect that is driven by the topology of the crystalline phase rather than by an attractive interaction like in Ref.~\cite{wunschd11}. 

\begin{figure}
\begin{center}
\includegraphics[width=0.45\textwidth]{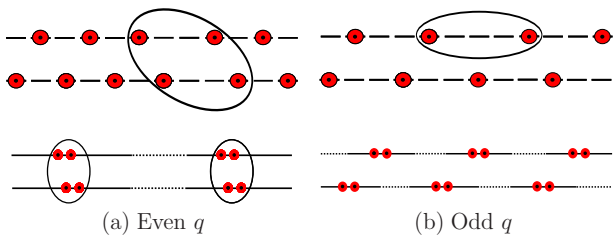}
\caption{(Color online) 
The configuration of dipolar bosons in 
the hole-like soliton states for fillings $p/q =1/2$ (a) and $p/q = 1/3$ (b), which have even and odd $q$, respectively. 
Top: configuration 
around one defect. Bottom: the arrangements of the defects along the tubes.
\label{fig:schematic}}
\end{center}
\end{figure}

We consider the following extended Bose 
Hubbard Hamiltonian for two tubes:
\begin{align}\notag
H &= - t\sum_{i,\alpha}\left( c_{i+1,\alpha}^{\dagger} c_{i,\alpha} + {\rm h.c.} \right) 
+ \sum_{i>j,\alpha} V(r_{ij}) \ \hat{n}_{i,\alpha} \hat{n}_{j,\alpha} 
\\ \label{eq:1}
& + \sum_{i,j} V\left(\sqrt{r_{ij}^2+d^2}\right) \ \hat{n}_{i,1} \hat{n}_{j,2} 
- \mu \sum_{i,\alpha} \hat{n}_{i,\alpha}  
\end{align}
The index $\alpha = \{1,2\}$ denotes the two different tubes,
 $\hat{n}_{i,\alpha} = c_{i,\alpha}^\dag c_{i,\alpha}$ gives the density at site $i$ on tube $\alpha$, $t$ is the hopping between neighboring sites,  and $V(r) = V_0/r^3$ is the dipole-dipole interaction with 
$V_0$ proportional 
to the square of the dipole moment.
The second and third terms in Eq.~\eqref{eq:1} correspond to the intratube and intertube interactions, respectively. 
The interparticle distances are parameterised in terms of the horizontal
distance $r_{ij} = |i-j|$, where we have set the lattice spacing in each tube to be 1,
and the intertube distance $d$ (see Fig.~\ref{fig:setup}). 
The on-site repulsion is taken to be infinite, so that we have hard-core bosons (or, equivalently, spinless fermions~\cite{giamarchi_book}). 
Note that the chemical potential $\mu$ is the same for each tube, corresponding to an equal filling fraction $p/q$ in each tube.
We consider filling fractions $p/q <\frac{1}{2}$
so that the total energy of the system remains finite in the limit $d \to 0$. 
However, we expect our main conclusions to also hold for $\frac{1}{2} \leq p/q < 1$.

We begin by considering the classical case of zero hopping ($t=0$). 
In the limit $d \to \infty$, we clearly have two isolated tubes and thus we recover the classical commensurate ground state (CGS)  for a single tube~\cite{hubbard78,pokrovskyu78}.
Here, we have a crystal commensurate with the lattice: for density $p/q$  in each tube, the configuration of particles has period $q$ with $p$ particles arranged in each period (assuming that $p$ and $q$ have no common factors). 
The configuration for a single tube 
can be determined using an algorithm proposed by Hubbard~\cite{hubbard78}. 
If we denote the distance between particles 1 and 2 as $x_1$, the distance between particles 2 and 3 as $x_2$ and so on, then the CGS configuration satisfies $|x_i - x_j| \leq 1$ for all pairs $i$, $j$.   
The same applies to 2nd neighbor distances $y_1 = x_1 + x_2$, $y_2 = x_2 + x_3$ etc.,  where we have $|y_i - y_j| \leq 1$, and so on for higher order $k$-th neighbor distances. 
This yields, for example, the configuration in Fig.~\ref{fig:setup}(b) for filling fraction 3/4. For Hubbard's algorithm to apply, it is sufficient to assume that the interaction potential is convex, i.e.\ $V(r+1) + V(r-1) \geq 2 V(r)$, where $r$ can correspond to any $k$-th neighbor distance.  

To determine the ground state for the two-tube system, we first take $d \to 0$ so that all the particles are effectively collapsed onto one tube. We then obtain the CGS for a single tube with filling fraction twice that of each tube, i.e.\ $2 p/q$. When the two tubes are drawn apart slightly with $d \ll 1$, the particle configuration is unperturbed (the CGS has a finite energy gap) and 
the repulsion is clearly minimized if we assign every second particle to each tube (see Fig.~\ref{fig:setup}).
This means that the configuration in each tube corresponds to the single-tube CGS for filling $p/q$, since each tube only contains the sets of 2nd, 4th, ..., $2k$-th neighbor distances of the collapsed tube,
all of which satisfy the CGS constraint discussed above.
Increasing $d$ further, Hubbard's argument ensures that this configuration in the two tubes remains the ground state as long as the intertube potential $V\left(\sqrt{r_{ij}^2+d^2}\right)$ remains convex with respect to $r_{ij}$. 
From an inspection of $\partial^2_r V\left(\sqrt{r^2+d^2}\right)$, we expect that convexity could be violated when $d>2 r_{ij}$. This could lead to a scenario where the crystal patterns in each tube become shifted with respect to one another once $d \gtrsim \text{min}(2r_{ij})$. However, we find that this shift does not occur for  
infinitely long tubes because of the terms in the energy involving large $k$-th neighbor distances where convexity is not violated. 
Thus, the CGS configuration remains the same for arbitrary $d$. 
Note, though, that this convexity violation does have consequences for 
the defect state discussed below.

To assess the stability of the two-tube CGS for each filling, we must consider adding/removing one particle to/from each tube.
For the single-tube case, such an addition (removal) leads to the formation of solitons or defects
in the tube pattern.
Indeed, for filling fraction $p/q$, the added particle or hole fractionalizes into 
$q$ solitons that are particle- or hole-like, respectively. 
For simple filling fractions $1/q$, the $q$ soliton states (qSS) involve solitons
corresponding to an arrangement where the length of the unit cell
is reduced (increased) by one vacant site~\cite{hubbard78,BakBruinsma}. For more general filling fractions,
the soliton has a more complicated structure, which can be derived from Hubbard's algorithm~\cite{hubbard78}. 
%

Returning to the two-tube case, we determine the qSS by starting from the collapsed tube ($d=0$) once again. By removing two particles (one from each tube) from the CGS for filling fraction $2p/q$, we immediately obtain two different scenarios: for odd $q$, we generate  $2 \times q$ solitons, while for even $q$, we have $2 \times q/2 = q$ solitons since the crystal period in this case is $q/2$. Now, when we separate the tubes, keeping $d \ll 1$, we obtain the single-tube hole-qSS for filling fraction $p/q$ on each tube, which gives $q$ solitons on each tube. However, the difference in topology between odd and even $q$ means that
for even $q$, these solitonic defects are bound in pairs across the tubes, while for odd $q$, all the solitons 
repel each other and 
are maximally spread out --- see Fig.~\ref{fig:schematic}(a) and (b), respectively. 
The same situation applies to the particle-like solitons generated when a particle is added to each tube. 

\begin{figure}
\begin{center}
\includegraphics[width = 0.4\textwidth]{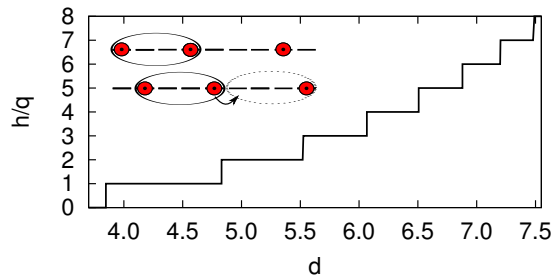}
\caption{(Color online) The separation $h$ between bound particle-like solitons as a function of intertube distance $d$ for $p/q=1/4$ on each tube. The solitons move apart in steps of $q=4$ and we define $h=0$ to be the maximally-bound configuration (inset).  The solitons start separating once $d\gtrsim 3.8$.
 \label{fig:sol_sep}}
\end{center}
\end{figure}

The pairs of solitons for even $q$ are ``maximally bound'' as in Fig.~\ref{fig:schematic}(a) as long as the configuration corresponding to the qSS for the collapsed tube remains the lowest energy state. However, as discussed earlier, this is not guaranteed once $d > \text{min}(2r_{ij})$ and indeed we find that soliton pairs start to separate above a critical $d$, becoming more weakly bound as $d$ is increased (see Fig.~\ref{fig:sol_sep}). 
The solitons eventually unbind in the  limit $d \to \infty$, where we recover the single-tube case. 

Following Ref.~\cite{BakBruinsma}, we determine the interval in chemical potential $\Delta\mu$ over which a given CGS is stable by comparing the CGS energy with the energies of the particle-qSS and the hole-qSS.
Like the single-tube case, we find that the two-tube system in the thermodynamic limit exhibits a complete devil's staircase, where every tube filling fraction $p/q$ enjoys a region of stability $\Delta \mu$ which all add up to fully cover the range of $\mu$. 
However, we also find that the energy gap $\Delta\mu$ depends on the intertube distance $d$ when $q$ is even, but not when $q$ is odd, as follows: 
\begin{align} 
\Delta \mu_{\rm odd} = &   
\sum_{m=1} mq F(mq,0)  \nonumber\\
\Delta \mu_{\rm even} = &  
\sum_{ m=1} mq F(mq,0) + \frac{(2m-1)q}{2} F\Big(\frac{(2m-1)q}{2},d\Big)  \nonumber\\
&+ \sum_{l=1}^{h} \sum_{m=1}^{l-1}  F\Big(\frac{(2m-1)q}{2},d\Big) \textrm{,}
 \nonumber
\end{align}
where $F(a,d)  = -2V(\sqrt{a^2+d^2}) + V(\sqrt{(a-1)^2+d^2}) +V(\sqrt{(a+1)^2+d^2}) $.
Thus, for large $q$, $\Delta \mu_{\rm odd}$ scales approximately as $1/q^4$ for all $d$, while $\Delta \mu_{\rm even}$ scales as $(2/q)^4$ when $d\to 0$ and as $1/q^4$ when $d\to\infty$.
This is consistent with the odd-even behavior of the solitons in the two-tube system. 

\begin{figure*}
\begin{center}
\includegraphics[width = 0.8\textwidth]{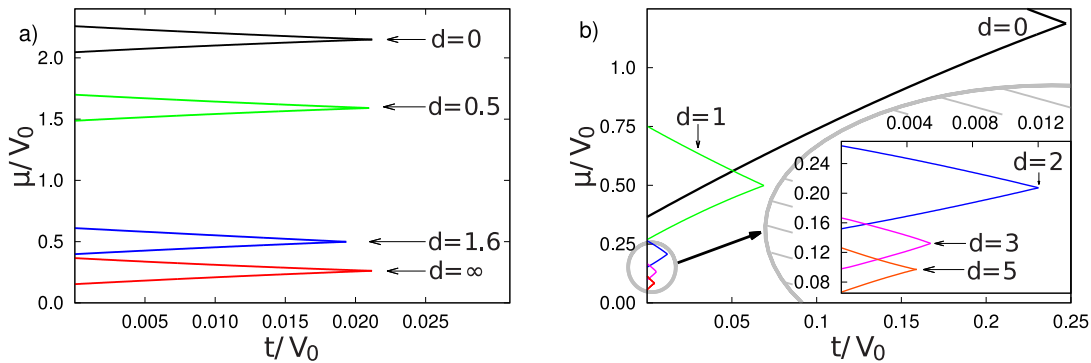}
\caption{(Color online) Mott lobes in the $\mu$-$t$ phase diagram corresponding to tube filling fractions (a) $p/q = 1/3$ and (b) $p/q = 1/4$ for different values of intratube distance $d$.
In the limit $d \to 0$, the lobes for $p/q=1/4$ and $p/q=1/3$ become equivalent to the lobes for a \emph{single} tube with filling fraction $1/2$ and $2/3$, respectively. 
The inset in (b) zooms into the Mott lobes for larger distances. 
\label{fig:tipmove}}
\end{center}
\end{figure*}

%
We now turn to 
 the case of finite hopping $t>0$. 
Here, the solitonic excitations can melt the CGS Mott phase into a Luttinger liquid~\cite{burnellpcs09}, leading to a series of Mott lobes reminiscent of the phase diagram for the Bose Hubbard model~\cite{fisherwgf89}. 
The phase boundary of each Mott lobe in the $\mu$-$t$ phase diagram (Fig.~\ref{fig:tipmove}) corresponds to when the energies of the CGS and its adjacent qSS are equal, i.e., for each $t$ and $d$, we take $E_{CGS}(\mu) = E_{qSS}(\mu)$~\footnote{This assumes that the phase transition is continuous, which is reasonable since the interactions between isolated solitons (for odd $q$) or bound pairs of solitons (for even $q$) are repulsive.}. 
Similarly to Refs.~\cite{burnellpcs09,freericksm96}, we determine the energies using a strong-coupling expansion in $t/V_0$. 
We expand up to second order in $t/V_0$, which is sufficient to capture the basic shape of the Mott lobes for odd $q$ like in Fig.~\ref{fig:tipmove}(a), but is not always enough for even $q$ as we discuss below. 
This perturbative approach is also never accurate at the lobe tip, where the energy gap of the CGS goes to zero, but we are in any case more concerned with the sides of the Mott lobe, where the CGS is doped with particles (upper boundary) or holes (lower boundary). 
The strong-coupling expansion for the CGS energy possesses the standard form: 
the first-order correction is zero, while the second order correction is $E^{(2)}_{CGS}=-2t^2\sum_{i=1}^N \frac{1}{|\Delta E_i|}$, where $N$ is the number of particles and $\Delta E_i$ is the difference in energies for $t=0$ between the ground state and the excited state created by hopping the $i$-th particle.

The qSS is highly degenerate in the classical limit and thus the relevant states to consider in the perturbation theory are the momentum eigenstates of the soliton. 
For odd $q$, the solitons are unbound and thus the corrections to the qSS energy 
have a similar form to those for the single tube in Ref.~\cite{burnellpcs09}. 
In this case, the first-order correction per added particle/hole 
is $-2tq$, corresponding to solitons with zero
momentum (which lie at the bottom of the band).
Figure~\ref{fig:tipmove}(a) depicts the Mott lobes up to second order in $t/V_0$ for $p/q = 1/3$, representative of the behavior for odd $q$ as a function of $d$. 
The width of the lobes at $t=0$ corresponds to $\Delta\mu_{\rm odd}$. We see that the shape of the lobe changes very little with $d$, owing to the fact that the solitons have the same structure in the limits $d\to 0$ and $d\to \infty$.

For even $q$ and $d\ll 1$, the first-order correction to the qSS energy is instead $-tq$ per added particle/hole, since the solitons are now bound in pairs. Here, each maximally-bound pair is hopped by $q/2$ sites when a soliton in the pair is hopped by $q$ sites (see Fig.~\ref{fig:schematic}(a)).
The second-order correction for the qSS (with one particle/hole added to each tube) has the form
\begin{align}
 E^{(2)}_{qSS}  = -2q \frac{t^2}{\Delta E_{r_1,1,-}} -q t^2 \sum_{i=1}^{N/2q} 
\sum_\alpha
\sum_{\beta = \pm}  \frac{1}{\Delta E_{r_i,\alpha,\beta}}  \ .
\nonumber 
\end{align}
Like before, $\Delta E_{r_i,\alpha,\beta}$ is the potential energy cost for hopping a particle, where $r_i$ is the position of the hopped particle on tube $\alpha$ with respect to the soliton on its tube. 
The particle can be hopped towards ($+$) or away from ($-$) its soliton. 
Note that degenerate states, where $\Delta E_{r_i,\alpha,\beta}=0$, are excluded.
The first term of $E^{(2)}_{qSS}$ is independent of $\alpha$ and corresponds to the soliton pair propagating via an intermediate state where the soliton pair pulls apart.

Beyond a critical $d$, the solitonic pairs  
at $t=0$ start to separate and the first-order correction to the qSS energy can then be zero.
This suggests that our simple perturbative expansion is inadequate for large $d$ since we expect the first-order correction to be $-2tq$ per added particle/hole in the single-tube limit $d\to\infty$.
Indeed, we find that the energy difference between states with different soliton separations 
 (Fig.~\ref{fig:sol_sep}) 
rapidly approaches zero with increasing $d$ so that the second-order correction $E^{(2)}_{qSS}$ becomes sizeable even for relatively small $d$. 
This implies that we must include states $\ket{h}$ with different soliton separations $h$ in our strong-coupling approximation for the energy. We construct the effective Hamiltonian $H^{{\rm eff}}$ for the set of states $\{ \ket{h} \}$ by including any contributions of excited states not included in $\{ \ket{h} \}$ up to second order in $t/V_0$. For instance, the diagonal elements of $H^{{\rm eff}}$ will contain the potential energies for each $\ket{h}$ plus a $t^2$ term due to fluctuations of the underlying commensurate pattern. 
This approach effectively amounts to performing a Schrieffer-Wolff transformation~\cite{schriefferwolff, macdonaldgy88} on the original Hamiltonian \eqref{eq:1}. 
We then diagonalise $H^{{\rm eff}}$ to determine the energy of the qSS. In practice, we consider 
  $\{ \ket{h}\}$ up to a maximum separation $h_{\rm max}$, where $h_{\rm max}$ is large enough that the lowest eigenvalue of $H^{{\rm eff}}$ does not depend on $h_{\rm max}$.

Figure~\ref{fig:tipmove}(b) depicts the resulting Mott lobes for even $q$, where we consider $p/q = 1/4$ for various $d$.  
For $d>2$, we determine the lobes using $H^{\rm eff}$ from the Schrieffer-Wolff transformation. 
We see that the size of the Mott lobes changes dramatically with $d$ and is dominated by how the width $\Delta\mu_{\rm even}$ at $t=0$ depends on $d$.
Note, also, that the slope of the Mott boundaries near $t=0$ becomes steeper with increasing $d$, approaching $d\mu/dt = \mp 2q$ for top and bottom boundaries, respectively. 
The evolution of the Mott lobes with $d$ provides an unambiguous signature of the soliton pairing for even $q$.

In principle, one can determine the nature of the Mott transition by approaching it from the 
Luttinger liquid~\cite{giamarchi_book,sela2011,weimerb10}. 
For a single tube, the transition at the Mott lobe tip is of the Kosterlitz-Thouless type, while the transition everywhere else is described by a two-band model of quasiparticles which are gapped in the Mott phase~\cite{burnellpcs09}. Further work is required to ascertain how this scenario is affected by soliton pairing and the different soliton-soliton interactions in the two-tube case. In particular, for odd $q$, the solitons are required to alternate between the two tubes, as depicted in Fig.~\ref{fig:schematic}(b), and thus the soliton-soliton interactions have an extra topological constraint that can impact the liquid phase.



Our predicted phase diagram should be experimentally realizable with ultracold dipolar atoms or molecules. 
For a typical lattice spacing of $500$nm within each tube and tunable lattice depths of $5E_r$ to $30E_r$, where $E_r$ is the recoil energy, the hopping $t$ can range from $0.1E_r$ to $5\times 10^{-4}E_r$. 
Thus, for $^{40}$K$^{87}$Rb polar molecules with dipole moment $\sim 0.2$ Debye as in current experiment~\cite{miranda2011}, we obtain $t/V_0 \simeq 0.02-3$, which is sufficient to observe the behavior of the Mott lobes in Fig.~\ref{fig:tipmove}(b). 
To enhance $V_0$, one can consider Rydberg-dressed atoms, where the effective dipole moment can reach $10$ Debye. This allows one to lower $t/V_0$ to 
well within the Mott lobes of Fig.~\ref{fig:tipmove}. 
The commensurate Mott phases could be probed by
Bragg spectroscopy~\cite{stoferle2004} and, for the case of Rydberg-dressed atoms,  
the solitonic pairs could be detected by single-atom-resolved fluorescence imaging~\cite{endres2011, shersonkuhr2010,bakrgreiner10}. 
There is also the prospect of probing the compressibility of the Mott phases locally in the trap now that single sites can be manipulated independently~\cite{weitenbergkuhr2011}.

We finally note that our two-tube system is also potentially connected to quantum Hall bilayers since the strong-coupling limit of each 1D lattice corresponds to the Tao-Thouless limit of the fractional quantum Hall effect~\cite{bergholtz2008,larson2012}. 

\acknowledgments We are grateful to R. Brierley, F. Burnell, M. Cheneau, M. Garst  and S. Baur for
useful discussions. 
MB acknowledges support from the Gates Cambridge Trust. 
MMP acknowledges support from the EPSRC under Grant No.\ EP/H00369X/1. 


\end{document}